\documentclass{emulateapj} 

\def\msun{{\rm M$_{\odot}$}}

\begin{document}

\title{Tracing the Dynamical History of the Globular Cluster 47~Tucanae}

\author {Eric Monkman \& Alison Sills}
\affil{Department of Physics and Astronomy, McMaster University, 1280
Main Street West, Hamilton, Ontario, L8S 4M1, Canada}
\email{asills@mcmaster.ca, monkmae@mcmaster.ca}

\author{Justin Howell}
\affil{Infrared Processing and Analysis Center,
Mail Stop 100-22,
California Institute of Technology,
Jet Propulsion Laboratory,
Pasadena, CA 91125}
\email{jhhowell@ipac.caltech.edu}

\author{Puragra Guhathakurta}
\affil{ UCO/Lick Observatory,
University of California,
271 Interdisciplinary Sciences Building,
1156 High Street,
Santa Cruz, CA 95064}
\email{raja@ucolick.org}

\author{Francesca de Angeli}
\affil{Institute of Astronomy,
University of Cambridge,
Madingley Road,
Cambridge CB3 0HA
United Kingdom}
\email{fda@ast.cam.ac.uk}

\author{Giacomo Beccari} 
\affil{INAF-Osservatorio Astronomico di Bologna,
via Ranzani 1, I-40127 Bologna, Italy;
Dipartimento di Scienze della Comunicazione,
Universita degli Studi di Teramo,
I-64100 Teramo, Italy;
INAF-Osservatorio Astronomico di Collurania,
Via Mentore Maggini, I-64100 Teramo, Italy}
\email{giacomo.beccari@bo.astro.it}

\begin{abstract}

We use two stellar populations in the globular cluster 47~Tucanae to
trace its dynamical history: blue stragglers and low mass main
sequence stars.  We assumed that the blue stragglers were formed
through stellar collisions in all regions of the cluster.  We find
that in the core of the cluster, models of collisional blue stragglers
agree well with the observations as long as blue stragglers are still
continuing to form and the mass function in the cluster is
extremely biased towards massive stars ($x=-8$ where a Salpeter mass
function has $x=+1.35$). We show that such an extreme mass function is
supported by direct measurements of the luminosity function of main
sequence stars in the centre of the cluster. In the middle region of
our dataset ($25''$ to $130''$ from the cluster centre), blue
straggler formation seems to have stopped about half a Gyr ago.  In
the outskirts of the cluster, our models are least successful at
reproducing the blue straggler data. Taken at face value, they
indicate that blue straggler formation has been insignificant over the
past billion years, and that a Salpeter mass function
applies. However, it is more likely that the dominant formation
mechanism in this part of the cluster is not the collisional one, and
that our models are not appropriate for this region of
the cluster. We conclude that blue stragglers can be used as tracers
of dynamics in globular clusters, despite our incomplete understanding
of how and where they were formed.
\end{abstract}


\section{Introduction}

Globular clusters in our Galaxy are often called the ideal
laboratories for a variety of fields in stellar astrophysics. They
provide the closest thing to controlled experiments in which to test
theories of stellar evolution and stellar dynamics. From the evolution
point of view, globular clusters provide a large population of stars
that have a common distance, metallicity, and age. Under the
assumption of single-star, undisturbed evolution, researchers have
used globular clusters to test models of star formation and evolution
of low-mass stars for decades. From the dynamics point of view,
globular clusters lie in a very interesting region of phase space for
stellar systems. Unlike open clusters, their crossing timescale is
longer than their relaxation timescale, meaning that they do not
dissolve quickly. Unlike galaxies, globular clusters have relaxation
times shorter than their evolution time, meaning that they are
dynamically evolved. Under the assumption of collisionless dynamical
evolution, globular clusters are very nice real-life examples of large
$N$ Newtonian gravitation systems.

If one looks at globular clusters very closely, their use as
laboratories for stellar evolution and stellar dynamics becomes less
obvious. The assumptions mentioned in the previous paragraph break
down. Not all stars in globular clusters have evolved in isolation;
not all stellar interactions in globular clusters are
collisionless. Our laboratory is now a place to study the interplay
between stellar dynamics and stellar evolution, and to look carefully
at the feedback between these two fundamental fields of stellar
astronomy.

In this paper, we attempt to use the effects of strong stellar
encounters, particularly stellar collisions, to probe the dynamical
history of the globular cluster 47~Tucanae. We will use two tracer
populations to do this. The first is blue stragglers,
core-hydrogen-burning stars that are bluer and brighter than the main
sequence turnoff. The second population is the low-mass main sequence
stars, specifically their mass function.

The effects of stellar collisions inside of dense globular clusters
play a significant role in the evolution of these systems
\citep{L89}. These collisions are one of a number of suspected
mechanisms for producing blue stragglers. Many other
mechanisms for the formation of these stars have been proposed, but
the most convincing ideas involve close, disruptive encounters: direct
collisions, or interactions within binary systems.

In a close binary, as the more massive star begins to swell on its
path to becoming a red giant, some of its material may be pulled into
the less massive star. This less massive star would then grow and
become a blue straggler \citep{MC64}. Close binary systems can also
gradually lose angular momentum, forcing the two stars to combine into
a blue straggler \citep{ZS76}. Most blue stragglers are more massive
than the turnoff mass, but less than twice it, and these mechanisms
are both able to account for these stars. However, both of these
methods are unable to explain the presence of blue stragglers with
masses greater than twice the turnoff mass, and this poses a
significant problem for these explanations \citep{L92}. Collisionally
formed blue stragglers do not suffer from this problem, as it is
possible, although rare, for three stars to collide, usually in an
interaction involving a binary star system \citep{L89}. Direct
measurements of masses of individual blue stragglers in globular
clusters are rare, but the work of \citet{deM05} suggests that $\sim 10$\%
of blue stragglers in globular clusters could be the result of triple
collisions. Three of their 24 blue stragglers, with masses measured
from {\it HST\/} STIS spectroscopy, have masses above 1.8
\msun. Interestingly, all 3 are within a few arcseconds of their
cluster centres.
 
The main problem with the collisional mechanism is that it has been
unable to explain all of the blue stragglers in any cluster. In
particular, the collisional probability for a given star drops to near
zero quite quickly outside of the core. For this reason, it has been
suggested that blue stragglers are formed through both binary and
collisional mechanisms \citep{M04,D04}. This suggestion is further
supported by the findings of \citet{P04}, who have shown that there is
no correlation between expected collision rate and blue straggler
frequency, and only minor correlations between this frequency and
cluster mass and central density. As a cluster becomes more dense,
stellar collisions become more likely, and blue stragglers formed
through this method increase in number. The increased density,
however, makes binary systems evolve faster due to exchanges of more
massive stars in the binaries during dynamical interactions.  If the
rise in one mechanism's efficiency is similar in magnitude to the drop
in another's, then this explains the lack of correlation between
expected collision rate and blue straggler frequency \citep{D04}.
 
In this paper, we examine the distribution of collisionally formed
blue stragglers in the colour-magnitude diagram. Using {\it Hubble
Space Telescope\/} ({\it HST\/}) and ground based observations of the
globular cluster 47~Tuc, we are able to study blue stragglers in
various regions of the cluster. We run simulations of the evolution of
collisional blue stragglers, varying the mass function of their
progenitor stars, as well as their formation times. We simulate three
regions of the cluster, separated from each other by radial distance
from the cluster centre, and find different formation times and
progenitor mass functions for each.  We discuss our data sets in
section 2. In section 3, we give the details of our simulations. We
show our results for each region of the cluster in section 4. Section
5 discuses mass segregation, and we summarize and discuss our results
in section 6.

\section{The Data}

We use three sources of data to get as much coverage of 47~Tuc's
stellar populations as possible. Two of these sets are taken from {\it
HST\/} images that cover the central region of the cluster. The third
is ground-based data that covers the rest of the cluster, out to
$1200''$ from the centre.

There are two {\it HST\/} imaging data sets analyzed in this paper:
({\bf 1})~archival Wide Field Planetary Camera~2 (WFPC2) data from
program GO-6095 (PI: S.~G.~Djorgovski), consisting of F218W
($4\times800$\,s), F439W ($2\times50$\,s), and F555W (7\,s) images;
hereafter the ``archival'' data set; ({\bf 2})~ultra-deep WFPC2 data
from \citet{gilliland}, consisting of $10^5$~s integration in each of
the F555W and F814W bands.  Sub-pixel dithering yields a point spread
function whose full width at half maximum is approximately $0.07''$
on the PC1 CCD.  Hereafter this will be referred to as the
``ultra-deep'' data set.  An analysis of the archival and ultra-deep
data sets is presented in the companion paper by \citet{guh06}.

The blue stragglers from the archival dataset were selected from the
data published in \citet{P02} and \citet{P04}.  We used the $B$ and
$V$ magnitudes of each blue straggler (calculated from F439W and F555W
magnitudes).  We needed to know the absolute position of each blue
straggler in RA and DEC, in order to combine this data set with the
ground-based data.  We used the STSDAS task `metric' to convert from
{\it HST} WPCS2 CCD coordinates. These data reach out to $130''$ from
the cluster centre.

Our wide field data are taken from \citet{F04}, and cover the cluster
out to $1200''$ from the centre. These data were taken using the Wide
Field Imager on the 2.2 m ESO-MPI telescope at La Silla, and we have
$B$, $V$, and $I$ magnitude information for each blue straggler. The
HST image is more suited to analyzing the dense central region than
the ground-based data. For this reason, the wide field data inside of
$130''$ is not given in \citet{F04}, and is not used in our analysis.

\begin{figure}
\plotone{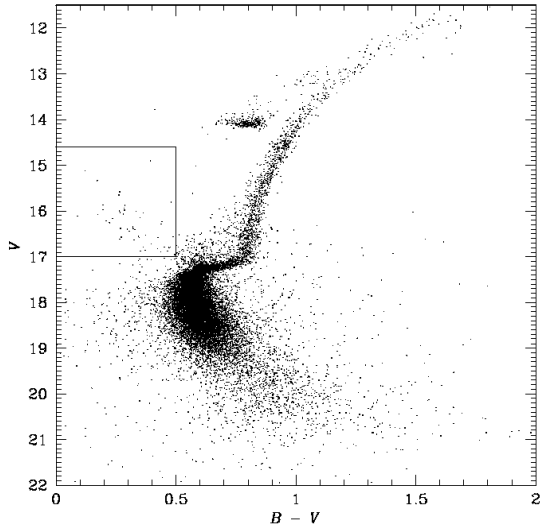}
\caption{The colour-magnitude diagram for 47~Tuc from the archival 
{\it HST\/}/WFPC2 data. The blue straggler selection box is
marked. \label{fullCMD}}
\end{figure}

The colour-magnitude diagram for 47~Tuc, taken from the archival WFPC2
data, is shown in Figure~\ref{fullCMD}. The blue straggler selection
box is shown as a solid line. This box encompasses the selection
criteria used on both the {\it HST} and ground-based data.

\begin{figure}
\plotone{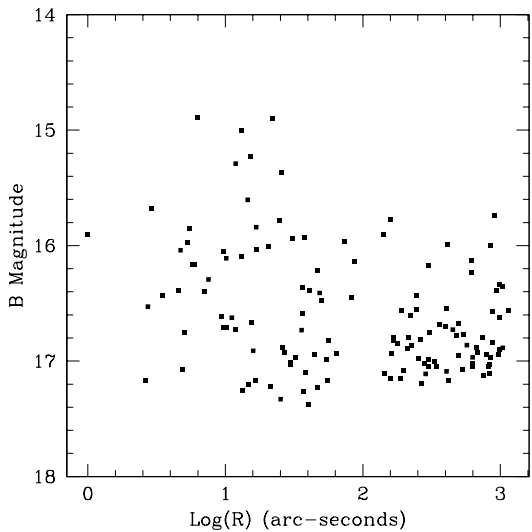}
\caption{Apparent $B$ magnitude versus radial 
distance from the centre of the cluster, in the logarithm of
arcseconds, for blue stragglers stars in our 47~Tuc sample. The number
of bright blue stragglers dramatically declines outside the cluster
core. \label{BvsR}}
\end{figure}

\begin{figure}
\plotone{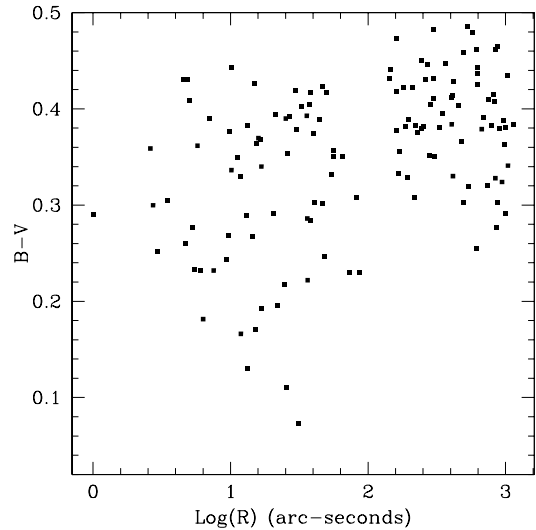}
\caption{The $B-V$ colour of 47~Tuc blue stragglers versus radial 
distance from the centre of the cluster, in the logarithm of arc
seconds. Notice the lack of bluer blue stragglers at large distances
from the cluster core.\label{CvsR}}
\end{figure}

We use the data from the archival and wide-field sources to plot both
magnitude and colour against radial distance from the centre of the
cluster, shown in Figures~\ref{BvsR} and \ref{CvsR} respectively. The
centre of the cluster is taken to be at Right Ascension
$\rm00^h24^m05.20^s$ and Declination $-72^\circ$04$'$51.00$''$. The
plot of magnitude against radial distance is in the B filter, colour
against radial distance is in B-V, and the radial distance for both is
given in the log of the distance in arc seconds. In both figures,
there is an empty region from approximately $100''$ to $130''$. The
shape of the HST field of view allows it to only cover a fraction of
this area \citep{F04}, and so no blue stragglers have been observed in
this region.

In the magnitude plot, there is clearly a drop in the number of bright
blue stragglers in the outskirts of the cluster. To determine the most
statistically significant value of the radius at which this drop
occurs, we used a Kolmogorov-Smirnov test to determine if the
populations inside and outside of a given value of $r$ were drawn from
the same parent population. We varied r from 1 to 150 arc seconds, in
steps of 1 arc second. We found that the most statistically
significant division is at $25''$, within the error of the
$24''\pm1.9''$ core radius of 47~Tuc \citep*{HGG00}. The KS test
returned a probability of more than 99.995\% that the populations to
the left and the right of the $25''$ line were drawn from different
distributions.  Bright blue stragglers are virtually non-existent
outside of this region. A similar pattern is also apparent in the plot
of colour against radial distance, indicating that the blue stragglers
in the outer region of the cluster are cooler than those of the inner
region.

\begin{figure}
\plotone{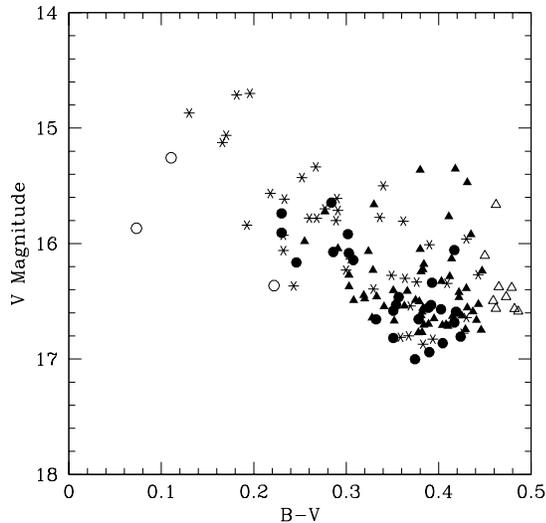}
\caption{Colour-Magnitude diagram for all of the blue stragglers in 
our 47~Tuc sample. The core blue stragglers, between $0''$ and $25''$,
are shown as asterisks; the blue stragglers that lie between $25''$
and $130''$ are shown as circles; and the outermost blue stragglers,
between $130''$ and $1200''$ are shown as triangles. The open symbols
highlight blue stragglers that pose problems for our simple
collisional models, and will be discussed in detail in section
\ref{Disc}.  \label{CMD}}
\end{figure}

We separate the data sets into three groups by radial distance from
the centre of 47~Tuc. The colour-magnitude diagram of all the blue
stragglers is shown in figure \ref{CMD}. The core of the cluster,
consisting of stars from $0''$ to $25''$, contains the brightest and
hottest stars and is shown as asterisks. The stars between $25''$ and
$130''$, shown as circles, encompass the rest of the {\it HST\/}/WFPC2
coverage. The ground-based field of view, from $130''$ to $1200''$,
contains none of the brighter or hotter stars, and is plotted as
triangles. The core, middle region of our dataset, and ground-based
field contain 44, 34, and 67 stars respectively. The open circles are
blue stragglers from the middle region of our data that are bluer than
any of our models, and the open triangles are blue stragglers that
posed problems for our models in the outermost region of our
data. Both populations will be discussed in detail in section
\ref{Disc}.

\section{Simulations}\label{Sims}
 
By assuming that all blue stragglers are created from single-binary
collisions, where two of the stars involved merge, and using the
method described in \citet{SB99}, we generate theoretical
colour-magnitude diagrams of the blue straggler distribution in
47~Tuc. The probability of the colliding single star having a certain
mass is taken from a power law mass function, representing the current
mass function in the region of interest. This same mass function is
used to find the probability of a binary system having a particular
total mass. The mass of the secondary component of the binary is
governed by a separate mass function, taken to have the Salpeter value
of +1.35 as its index, which represents the mass function of the
cluster when the binary was formed. All of these mass functions are of
the form \(dN/dM = M^{-(1+x)}\), where a Salpeter mass function has a
value $x=+1.35$. The STARLAB software package is used to find the
cross sections for binary-single collisions
\citep{MH96}.

Using the initial mass of the collision product, we interpolate an
evolutionary track from the nearest simulated tracks. These simulated
tracks are calculated using the Yale Rotational Stellar Evolution
Code, YREC \citep{GDKP92}. The initial conditions for the simulated
tracks are generated with a smooth particle hydrodynamics simulations
of colliding stars \citep{SB99}.

To produce a more accurate representation of the cluster, we are able
to specify the time frame over which these stellar collisions
occur. We alter the start and end times for a constant collision rate,
and assume that the collision rate remains constant during this period
as in \citet{S00}. This is still not an accurate representation of the
dynamics of a cluster, but it allows us to generate reasonable models,
and extract information on the approximate ages of the blue straggler
populations.  Finally, we weight each collision product evolutionary
track by its probability of occuring, and generate a predicted blue
straggler distribution in the colour-magnitude diagram.

The simulations require various values to be assumed. The velocity
dispersion is taken 11~km~s$^{-1}$ , and the
binary fraction is set at 10\%, within error of the predicted values
\citep{G95,A01}. Both of these values have very little impact on the
shape of the density function, and have a much larger effect on the
total number of blue stragglers predicted. The metallicity of the
cluster is more influential on the shape of the plot, and is set at
$\rm[Fe/H] = -0.66$ \citep{Gr03}. \citet{SB99} explore the effects of
changing these quantities in detail.

Due to mass segregation, each of the three cluster regions, $0''$ to
$25''$, $25''$ to $130''$, and $130''$ to $1200''$, could have very
different mass functions. To improve the accuracy of our models, we
simulate each region using a variety of mass function indices. We make
the unphysical but simple assumption that the mass function remains
constant in time over the entire simulation, and that, within each
region, it does not change with distance from the core. This
assumption is reasonable for clusters, like 47 Tuc, with short
relaxation times. Most mass segregation will occur early in the
cluster's lifetime, before our simulations begin. By varying
both the collision time frame, and mass function of the constituent
stars, we find the best model for each region.

\section{Results \label{RSLTS}}

\subsection{The Core of 47~Tucanae}

\begin{figure}
\plotone{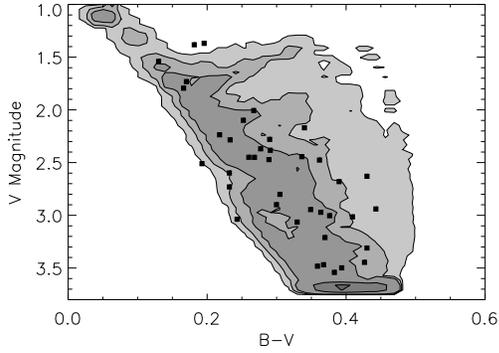}
\caption{A model of the blue stragglers in the core of 47~Tuc,
starting collisions 7~Gyr ago and ending them at the present day. This
model uses a mass function of $-8$.  The observed blue stragglers are
represented by squares. The contour levels are at 0.1\%, 0.5\%, 1\%,
5\% and 10\% of the total population. The models with blue straggler
formation starting at earlier than 7~Gyr ago were all very similar to
those shown here. \label{coresim}}
\end{figure}

As can be seen in Figure~\ref{coresim}, the innermost region of 47~Tuc
contains the brightest blue stragglers in the cluster. The
concentration of these stars is weighted much more toward the brighter
and bluer end of the diagram than either of the other two areas,
suggesting that the core population contains the most massive, and
youngest, blue stragglers.

To find an accurate model of the $0''$ to $25''$ region, we generate
colour-magnitude diagrams for a variety of collision time-frames. We
find that the models that best approximate this area are those that
start collisions at some time in the past and continue to the current
time. The brightest blue stragglers rapidly evolve out of the blue
straggler region of the colour-magnitude diagram, and stopping the
collisions a short time ago leaves this area void of any stars.

We test various start times for the collision process, and find that
the core of 47~Tuc has probably been producing blue stragglers for at
least the past 7~Gyr. Comparing the observed distribution of blue
stragglers with our models using a two-dimensional Kolmogorov-Smirnov
test, we find that at a start time of 7~Gyr ago, the best fitting
model is produced. More recent start times rapidly drop in their
ability to reproduce observations. Starting the collision process
further back than this produces models that are almost as closely
fitting to the data.

Upon establishing the time frame over which the collisions occur, we
test various mass functions and find the best-fitting slope. With a
mass function of $-8$ and with collisions occuring until the present
day, we find that by starting the collision process 7~Gyr ago, we can
reproduce the observed distribution of blue stragglers with a
certainty of 90\%. Starting the collisions between 8~Gyr ago and
12~Gyr ago produces a slightly better match at 92\%-93\%. This lack of
sensitivity on the older limit on the age reflects the low luminosity
cut-off of our blue straggler data, and is not an indication that old
blue stragglers do not exist. A mass function of $x=-8$ is a very
extreme power law, heavily weighted towards massive stars. As we will
show in section \ref{MF}, direct observations of the main sequence
stars in the cluster support such a strongly weighted mass function in
the core of 47 Tuc, although the exact value of the slope is not quite
the same ($x=-5$). We will argue in section \ref{Disc} that the result
from the direct observations should be taken as the true value, since
our blue straggler collision models include some simplifying
assumptions. However, the general agreement between these models and
the unrelated observations is extremely encouraging.

\subsection{The Middle Region of Our Dataset}

The $25''$ to $130''$ part of the cluster contains none of the
brightest blue stragglers, but still has a significant population of
moderately bright stars. This region has the smallest number of blue
stragglers out of the three regions we have chosen. Partly this is due
to the shape of the {\it HST\/}/WFPC2 chip, but it has also been shown
that blue stragglers are less prevalent in this area of 47~Tuc
\citep{F04}.  The middle region is clearly the transition between the
other two areas, being bluer than the outer region, and fainter than
the core. It is reasonable to expect that the time frame and mass
function that fit this region will also be a transition between the
other two. In this area, three stars cannot be accounted for by our
simulations. These stars are the only blue stragglers in the entire
cluster that appear to the left of all of our simulated tracks in the
colour-magnitude diagram. These three stars are shown as open circles
in Figure~\ref{CMD}, and are not included in our analysis of this
region. They are discussed in further detail in the Summary \&
Discussion section.

In varying the collision time frame for this region, we found that by
stopping the blue straggler formation process 0.6~Gyr ago, we were
able to most closely approximate the blue straggler distribution in
the colour-magnitude diagram.  We then varied the mass function for
this and similar timeframes, and found that the data is most
accurately represented using a mass function with slope -3.

\begin{figure}
\plotone{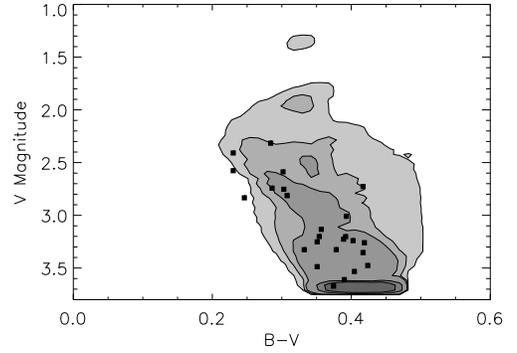}
\caption{ The colour-magnitude diagram of models of blue 
stragglers just outside the core of 47~Tuc. The model starts
collisions 7~Gyr ago, ends them 0.6~Gyr ago, and uses a mass function
slope of -3. The observed blue stragglers are represented by squares,
and the contour levels are the same as for
Figure~\ref{coresim}.\label{midsim}}
\end{figure}

Using this mass function and starting the collision process 7~Gyr ago,
we achieve the closest fit of 51\% probability that the data and model
were drawn from the same distribution, while starting it 8~Gyr ago
gives a similar fit of 49\%. A start time of 6~Gyr also works
reasonably well, at 44\%, but start times before or after these three
drop in effectiveness. The colour-magnitude diagram of the best-fit
model is given in Figure \ref{midsim}. If we include the three bluest
stars in our analysis, the KS probability for the best-fit model drops
to 27\%. These stars are discussed in detail in section \ref{Disc}.

\subsection{The Outermost Region}

The outer region of the cluster, between $130''$ and $1200''$ from the
cluster centre, contains a less diverse population than the other two
areas. As can be seen in Figure~\ref{CMD}, all of the stars in this
region are in the same group in the colour-magnitude diagram. The
colour of these stars suggests that they are older and less massive
than the other blue straggler populations. There is a dense group of
the coolest stars whose density is never accounted for by our
simulations. Our evolutionary tracks travel quickly through this
region, and so we predict a smaller number of blue stragglers than we
observe. We analyze this region with and without these 9 coolest
stars, to ensure that they do not mask any results for the rest of the
region. We discuss these 9 blue stragglers in further detail in the
Summary \& Discussion section.

We find that, whether we include the coolest stars or not, our
simulations best fit the data when we stop the collision process
1.2~Gyr ago. We adjust the start times of this process, and see that
the data is most accurately represented by a collision process that
starts between 9 and 7~Gyr ago. We are most successful at representing
this data when using a mass function with a slope of +1.35, the
Salpeter value. This supports a bias toward more low-mass stars in
this part of the cluster.

\begin{figure}
\plotone{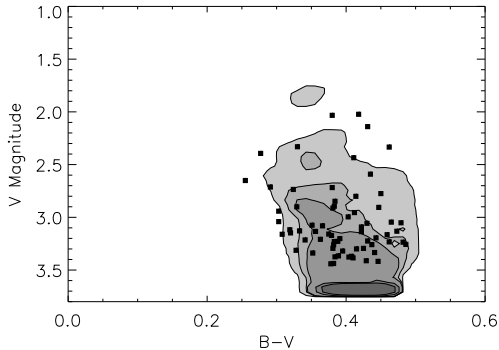}
\caption{Model of the blue stragglers that are farthest 
from the cluster centre. Observed blue stragglers are represented by
squares, and the contour levels are the same as for
Figure~\ref{coresim}. All shown models use a mass function with slope
of +1.35, start collisions 7~Gyr ago, and end collisions 1.2~Gyr
ago. \label{farsim}}
\end{figure}

When including the coolest stars in this area, we are not able to
produce convincing models of the data. We achieve our most accurate
model by starting the collision process 7 Gyr ago, producing a model
that matches the data to 17\%. When not including the coolest
stars, our models are much better at fitting the data. We use a
collision start time of 7 Gyr, with an end time of 1.2 Gyr, and find
that this models the data to within 35\%. Our best model can be
seen in Figure~\ref{farsim}.

These results indicate that the blue straggler population at the edge
of 47~Tuc is much older than in the rest of the cluster. We see that
the production of these blue stragglers must have begun around the
same time as in the rest of the cluster, but has stopped earlier. As
we are unable to accurately reproduce the data in the outermost region
of 47~Tuc using only the collision mechanism, our simulations support
binary mergers and mass accretions as being the dominant forms of blue
straggler formation in the outskirts of this cluster, in agreement
with \citet {M04}. These formation mechanisms would produce
distributions in the colour-magnitude diagram that are somewhat
similar to those produced through collisions, and so the small
agreement between our simulations and the data would be expected in
this case.

The basic problem with our models for this region of the cluster is
that the collision models and the assumptions we have made about the
dynamical state of the cluster do not accurately reproduce the balance
of faint and bright blue stragglers. We can find models that match the
brighter blue stragglers, but if we do, then we predict far too many
faint blue stragglers. Any mechanism that is at work in the outskirts
of the cluster must produce a more uniform distribution of blue
stragglers in luminosity than the collision models used here.

\section{Mass Segregation in the Core of 47~Tucanae \label{MF}}

The collisional model used to interpret our blue straggler data
predicts a mass function that is strongly weighted towards massive
stars at the centre of the cluster, and should become less top heavy
as one moves to the outskirts of the cluster.  Next we compare these
model predictions to the stellar mass function derived by comparing
the observed luminosity function in a succession of radial bins in the
central region of 47~Tuc to theoretical ones obtained from isochrones
based on a range of power-law mass functions \citep{vdb}.  The depth
and improved resolution of the ultra-deep data set result in complete
photometry fainter than $M_V=+8$ at all radii and are therefore well
suited for the measurement of the mass function.  The details of the
derivation of stellar mass function as a function of radius in 47~Tuc
are given in the companion paper by \citet{guh06}.

\begin{figure}
\plotone{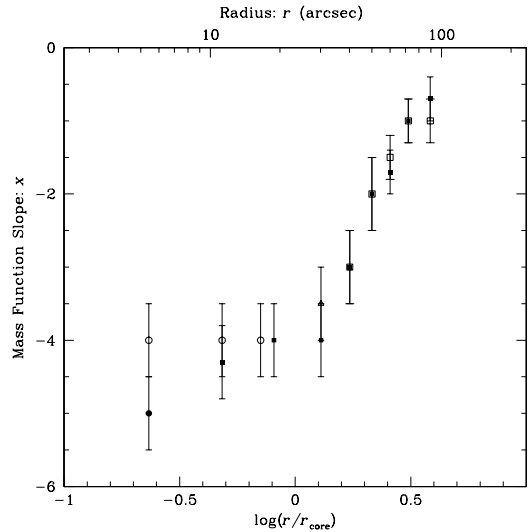}
\caption{The power-law mass function exponent $x$ in a series of 
radial bins.  Circles represent bins on the PC CCD, while squares are
on the WF CCDs.  Open points are from the archival data set, while
solid points are from the ultra-deep data set.  The open triangle is
the inner most radial bin on the WF CCDs of the archival data set,
which is most affected by stellar crowding.
\label{xr}}
\end{figure}

As shown in Figure~\ref{xr}, the stellar mass function slope in 47~Tuc
increases from $x=-5$ in the innermost bin to $x\sim-1$ in the
outermost radial bins.  It should be noted that \citet*{psm86} predict
mass function slopes increasing from $x=-3$ to $x\sim0$ over the same
range in radius in the most centrally concentrated King models
considered.  Thus, the extremely negative mass function in the core of
47~Tuc determined directly from the observations ($x\sim-5$) agrees
reasonably well with the value determined from the blue straggler
models ($x\sim-8$), and with dynamical predictions of mass
segregation. Given the uncertainties and simplifying assumptions
inherent in the blue straggler models, we argue that these two values
are in good agreement. The true mass function in the core may not
actually be as steep as $x=-8$, but two independent measures tell us
that it cannot be much flatter, and it certainly is not anything like
a Salpeter or other normal initial mass function. Our blue stragglers
models with a mass function of $x=-5$ fit the observations in the
inner region at the 67\% level, which is pretty good.  We should also
note that the blue straggler models for the 25'' to 130'' region of
the cluster also favour a mass function that agrees with the observed
value of $x\sim-3$.

\section{Summary \& Discussion}\label{Disc}

We found that, using the collision mechanism for blue straggler
formation, we were able to reproduce the observed distribution of blue
stragglers in colour-magnitude diagrams, to varying precision in
different areas of 47~Tuc. We used simulations to find the time over
which these stars were produced, along with the mass function of the
progenitor stars. We separated our data on blue stragglers by dividing
the stars into three groups, categorized by radial distance from the
cluster centre, and reproduced each group separately.

In our outermost region, extending from $130''$ to $1200''$ from the
cluster centre, our models matched the data to a maximum accuracy of
35\%. In order to achieve this match, our simulations began the
formation of blue stragglers between 7 and 9~Gyr ago, and ended it
1.2~Gyr ago. We found that using a mass function index of +1.35, the
Salpeter value, produced the closest fit to the data. The middle
region of 47~Tuc, $25''$ to $130''$ from the cluster centre, had a
younger population of blue stragglers than this outermost area. We
found that by starting stellar collisions between 6 and 8~Gyr ago, and
ending them 0.6~Gyr ago, we could reproduce the data to a higher
degree of certainty than for the outer region. A mass function with
slope of -3 most accurately represented our data, giving a 50\%
probability that the data and model were drawn from the same
distribution. We rediscovered the cluster core, with a $25''$ radius,
based on the blue straggler photometric information alone. We started
the blue straggler formation at various times, from the birth of the
cluster to 8~Gyr ago, and let it continue until now.  We found that a
mass function slope of -8 allowed us to reproduce our data with an
over 90\% certainty.

In the outermost region of 47~Tuc, a group of the 9 coolest blue
stragglers were not well-represented by any of our simulations. To
ensure that these stars alone were not interfering with our results,
we analyzed the region with and without including them. In both cases
it was found that we were unable to account for the outermost
population of blue stragglers using only the collision mechanism, but
the nature of these 9 stars is still unknown. If they are collisional
in origin, then they would have to be the result of a very rapid burst
in formation. If there was a burst of collisions at the correct time,
involving stars of a particular mass, then all those stars would be at
the same place on their evolutionary track, producing the observed
enhancement in that particular region of the colour-magnitude diagram.
This scenario is extremely unlikely. A better solution is probably
that these stars are not collisional in origin, and it is then outside
of the scope of this paper to predict how and when these blue
stragglers were produced, although we can say that they must be
relatively old. Given their location in the Hertzsprung gap, other
scenarios must be considered as well, such as whether these stars are
actually blends, either real (i.e. binary systems) or due to
observational error. Such blue stragglers have been found in almost
every other globular cluster that has been observed carefully.

In the middle region of our dataset, three blue stragglers were not
included in our analysis. These three stars were hotter than any of
our evolutionary tracks at their respective luminosities, and so our
simulations were incapable of reproducing them. These stars were the
only ones in the cluster that had this difficulty, which indicates
that they are different in origin from the other blue stragglers, and
are almost certainly not collisional. The existence of these strange
blue stragglers lends further support to our conclusion that the blue
stragglers in this area of the cluster are not entirely collisional in
origin. What is needed now is evolutionary tracks of binary merger
products, so that the same kind of simulations as presented in this
paper can be done with a different model for blue straggler
evolution. These models have been completely lacking in the past; we
look with interest to the preliminary models presented by
\citet{TDHZ06}.

Both sets of blue stragglers that are not well-represented by our
models lie reasonably close to the boundaries of our models. It is
possible that photometric errors are responsible for the seeming
disagreement with the models. However, in order for that to be the
case, the photometric errors on these stars must be high, and the
errors must be biased so as to move all the stars towards the models. We
find it unlikely that such a fortunate coincidence has occurred.

While there has almost certainly been some rate of stellar collisions
throughout the life of 47~Tuc, we have shown that the majority of
these collisions have been occuring since early in the life of the
cluster, but outside of the core the collision rate has substantially
dropped within the past few hundred million years. We found that the
regions that are farthest from the core have not had a significant
collision rate for the longest amount of time, and that as we look
closer to the centre, we see more and more evidence of recent
collisions. We showed that the farthest region of 47~Tuc has probably
not had a high rate of blue straggler formation for over the past
\(\approx\)1 Gyr, and that the region close to the core has not been
producing many blue stragglers for about half as long. The
distributions depend much less on the start times than on the end
times, so we were unable to isolate specific start times for our
collisions. When the older stars evolve over such a large period of
time, most of them become white dwarfs and leave the observable range
of objects. These stars will then have no bearing on the final
distribution of blue stragglers, and so the majority of the stars
produced from an earlier start time do not affect our results. The end
times that we found are reasonable, as one would expect stellar
collisions to still be a significant part of the dynamics in the very
dense core of a globular cluster. It also seems reasonable that in the
further out regions, where densities are very low, there would not be
a very high rate of blue straggler production.

We showed that the stars involved in blue straggler formation in the
outermost region of 47~Tuc were described by a mass function with
slope +1.35. This mass function favours lower mass stars over higher
mass ones, and indicates that in the outer region, where one would
expect the mass function of all the stars to have a value similar to
this, the blue stragglers are formed approximately equally by all
masses of stars. This result was echoed in our findings for the
intermediate region, where it was shown that a mass function index of
$-3$ produced the best fit to the data. In \S5 we showed that the mass
function near the outer edge of this region is $-2$, and then rises to
$-5$ in the centre of the core. The mass function of $-3$ is an
intermediate value in this region, and indicates that, like in the
outer region, the distribution of masses in the blue straggler
progenitors are representative of the overall stellar mass
distribution.

In the core of 47~Tuc, we found that a mass function slope of $-8$ fit
best with our data.  This low mass function index was partially caused
by our simulation, which assumed that all stars in the cluster were
traveling at the same velocity of 11~km~s$^{-1}$. If the stars in the
core of 47~Tuc all had similar kinetic energies, as they are expected
to through energy equipartition, then the square of their speeds would
be inversely proportional to their masses, and the more massive stars
will be moving slower than the less massive ones. The gravitational
cross section for a stellar collision is inversely proportional to the
relative velocity between the colliding stars, and so a collision
involving a slower moving star should have a larger gravitational
cross section than a collision involving faster moving stars. This
increase in the gravitationally focused cross section will raise the
collision rate for higher mass stars, beyond that which we took
account of. In order to compensate for this, our simulation has to use
a mass function that is more heavily weighted toward massive stars
than is actually be present, and so the very steep mass function that
we found is, in part, caused by our simulation's assumption of
constant stellar speed. This result is most noticeable in the core of
the cluster due to the much higher concentration of high mass
stars. Even with this effect, however, the progenitors of the blue
stragglers in the core of 47~Tuc were still significantly more massive
than in the rest of the cluster. The simulations we ran are not able
to account for this effect directly by changing the velocities of the
individual stars. However, if we simply reduce the velocity dispersion
in the cluster by a factor of 2 (to 5~km~s$^{-1}$) and perform our
simulations again, we find that our best fit scenario still has a mass
function of $x=-8$, but that it is only slightly better than the same
simultion with $x=-5$. It is also apparent the more massive blue
stragglers are slightly better fit with the higher mass function
index. that Therefore, we conclude that our mass function index
derived from the blue stragglers is indeed biased towards very low
values.

We found that it was much easier to replicate the distribution of blue
stragglers in the core than in the rest of the cluster. This suggests
that the blue stragglers in the core may be formed almost entirely
through stellar collisions. The number of collisions rapidly increases
with increasing density, and so, in the core of the cluster, where
densities are orders of magnitudes larger than in the outer regions,
stellar collisions should be a much more dominant effect. The rest of
the cluster should be less effected by collisions, and our results
show that outside of a few core radii, stellar collisions are much
less influential on the blue straggler distribution than other
formation mechanisms. This is in agreement with the predictions of
collision probability in the outskirts of the cluster. 

The models of blue straggler formation in the core of the cluster
suggest that the mass function is strongly weighted towards massive
stars, and the {\it HST\/} results confirm this for the present
day. However, the fact that all blue stragglers in the core require a
strongly negative mass function suggests that the mass function has
been this unusual for at least the lifetime of the brightest blue
stragglers, on the order of a few hundred million
years. Interestingly, the timescale for energy equipartition for most
globular clusters is also on the order of $10^8$ years. It may be that
detailed studies of the cores of other globular clusters will also
show this extreme mass function. If this is the case, we would expect
that the luminosity function of blue stragglers in the cores of
clusters should extend to brighter (i.e. more massive) stars than the
luminosity function in the outskirts of the cluster.

Blue stragglers are touted as tracers of dynamics in globular clusters
\citep{H93,SB99} since they are thought to form during stellar
encounters. Recent results from \citet{P04} have brought this claim
into question, since they find no correlation between blue
straggler populations and the expected collision rate in the
cluster. In this paper, we have shown that using the blue stragglers
alone, we can discover the core radius of 47~Tuc at least as well
as traditional methods. Or, to say it more accurately, the blue
stragglers inside and outside the core radius are drawn from clearly
different populations. We have also shown that we can find evidence of
mass segregation by applying a very simplistic model of blue straggler
formation to the blue straggler populations in different regions of
the cluster. Therefore, blue stragglers must have some connection to
globular cluster dynamics. The simple picture of blue straggler
formation (i.e. stellar collisions in globular clusters, merger of
primordial binaries in other environments) is too simplistic, but this
interesting population can still be used to study globular clusters in
their entirety.

In all regions of 47~Tuc, blue stragglers started forming at very
early times in the cluster -- at least 7~Gyr ago, although our data
are not sufficient to probe further back than that. However, in the
outskirts of the cluster, formation of bright blue stragglers ended
before it ended in the core. There are two possible reasons for this
(both of which could be operating). Either, bright blue stragglers can
only be formed in collisions, or the populations of binary stars are
different inside and outside the core. Collisions are the only way to
produce blue stragglers with masses significantly greater than twice
the turnoff mass, and therefore it may be reasonable to suggest that
the brightest blue stragglers are, in fact, triple collisions. If the
binary population inside the core consisted of a large number of
massive, hard binaries, then encounters between these binary systems
would result in more triple collisions than the encounters outside the
core. Binaries in the outskirts of the cluster would have to be either
less massive or consist of only wide binaries that could not merge in
the cluster lifetime. Is it possible that all the
blue-straggler-producing binaries could have migrated into the core
during the cluster lifetime? Or that binaries in the core are made to
be more massive through dynamical interactions \citep{D04}? The
timescales for mass segregation discussed above suggest that this is
likely, and the dynamical models of \citet{M04} are in agreement as
well (although in their model, some fraction of the binaries that can
produce blue straggler remain outside the core and produce the
peripheral blue stragglers). In both the Mapelli and Davies models, blue
stragglers were given a finite lifetime, but their position in the
colour-magnitude diagram was not considered. An obvious next step
would be to combine such models with blue straggler evolutionary
tracks, as an improvement over the simple models presented in this
paper.

There are an increasing number of recent studies of blue stragglers in
globular clusters that are combining ground-based wide-field data with
high quality HST data for the cores of clusters
\citep{F04,SFSR04}. The same kinds of models can be, and should be,
created for these clusters. We have very interesting indications that
something dynamical is happening to the blue straggler populations in
clusters with the evidence that the `unusual' bimodal population of
blue stragglers in M3 is not unusual at all, but seen in at least four
other clusters \citep[47~Tuc, M55 and NGC 6752; see][]{F06} and
\citep[M5; see][]{WSB06}). What does the distribution of blue
stragglers in the colour-magnitude diagram look like in the different
regions of these clusters? And how can we explain these results? We
suggest that blue stragglers do trace the dynamical state and history
of their parent globular clusters, and that their secrets can be
unlocked through detailed and consistent studies of their populations
in a variety of clusters.

\acknowledgements  EM thanks Alison Sills for her guidance and assistance
throughout this project. AS is supported by NSERC. EM was supported in
part by Human Resources and Development Canada.


\begin{thebibliography}

\bibitem[Albrow et~al.(2001)]{A01} Albrow, M.\ D., Gilliland, R.\ L.,
Brown, T.\ M., Edmonds, P.\ D., Guhathakurta, P., \& Sarajedini, A.\ 2001,
\apj, 559, 1060

\bibitem[Bergbusch \& VandenBerg(2001)]{vdb} Bergbusch, 
P.~A., \& VandenBerg, D.~A.\ 2001, \apj, 556, 322

\bibitem[Davies et~al.(2004)Davies, Piotto, \& De Angeli]{D04} Davies,
M.\ B., Piotto, G., \& De Angeli, F.\ 2004, MNRAS, 349, 129

\bibitem[De Marco et al.(2005)]{deM05} De Marco, O., Shara, 
M.~M., Zurek, D., Ouellette, J.~A., Lanz, T., Saffer, R.~A., \& Sepinsky, 
J.~F.\ 2005, \apj, 632, 894 

\bibitem[Ferraro et~al.(2001)]{F01} Ferraro, F.\ R., D'Amico, N.,
Possenti, A., Mignani, R.\ P., \& Paltrinieri, B. 2001, \apj, 561, 337

\bibitem[Ferraro et~al.(2004)]{F04} Ferraro, F.\ R., Beccari, G.,
Rood, R.\ T., Bellazzini, M., Sills, A., \& Sabbi, E. 2004, \apj, 603,
127

\bibitem[Ferraro(2006)]{F06} Ferraro, F.\ R. 2006, ASP Conf.~Ser.~TBA, 
Resolved Stellar Populations (astro-ph/0601217)

\bibitem[Gebhardt et~al.(1995)]{G95} Gebhardt, K., Pryor, C.,
Williams, T.\ B., \& Hesser, J.\ E. 1995, \aj, 110, 1686

\bibitem[Gilliland et~al.(2000)]{gilliland} Gilliland, R.~L., et~al.\ 2000,
\apjl, 545, L47

\bibitem[Gratton et~al.(2003)]{Gr03} Gratton, R.\ G., Bragaglia, A.,
Carretta, E., Clementini, G., Desidera, S., Grundahl, F., \&
Lucatello, S.\ 2003, \aap, 408, 529

\bibitem[Guenther et~al.(1992)]{GDKP92} Guenther, D.\ B., Demarque,
P., Kim, Y.-C., \& Pinsonneault, M.\ H.\ 1992, \apj, 387, 372

\bibitem[Guhathakurta et~al.(2006)]{guh06} Guhathakurta, P., Howell, J.\ H.,
Clem, J.\ L., Gilliland, R.\ L., \& Sills, A.\ 2006, \apjl, submitted
(astro-ph/0605XXX)

\bibitem[Howell et al.(2000) Howell, Guhathakurta, \&
Gilliland]{HGG00} Howell, J.\ H., Guhathakurta, P., \& Gilliland, R.\
L. 2000, \pasp, 112, 1200

\bibitem[Hut(1993)]{H93} Hut, P.\ 1993, ASP Conf.\ Ser.\ 53: 
Blue Stragglers, 53, 44
 
\bibitem[Leonard(1989)]{L89} Leonard, P.\ J.\ T. 1989, \aj, 98, 217

\bibitem[Leonard \& Linnell(1992)]{L92} Leonard, P.\ J.\ T., \&
Linnell, A.\ P. 1992, \aj, 103, 1928

\bibitem[Mapelli et~al.(2004)]{M04} Mapelli, M., Sigurdsson, S.,
Colpi, M., Ferraro, F.\ R., Possenti, A., Rood, R.\ T., Sills, A., \&
Beccari, G.\ 2004, \apjl, 605, L29

\bibitem[McCrea(1964)]{MC64} McCrea, W.\ H.\ 1964, \mnras, 128, 147

\bibitem[McMillan \& Hut(1996)]{MH96} McMillan, S.\ L.\ W., \& Hut,
P.\ 1996, \apj, 467, 348

\bibitem[Piotto et~al.(2002)]{P02} Piotto, G., King, I.\ R.,
Djorgovski, S.\ G., Sosin, C., Zoccali, M., Saviane, I., De Angeli,
F., Riello, M., Recio- Blanco, A., Rich, R.\ M., Meylan, G., \&
Renzini, A.\ 2002, \aap, 391, 945

\bibitem[Piotto et~al.(2004)]{P04} Piotto, G., De Angeli, F., King,
I.\ R., Djorgovski, S.\ G., Bono, G., Cassisi, S., Meylan, G.,
Recio-Blanco, A., Rich, R.\ M., \& Davies, M.\ B.\ 2004, \apjl, 604,
L109

\bibitem[Pryor et~al.(1986)Pryor, Smith, \& McClure]{psm86} Pryor, C.,
Smith, G.\ H., \& McClure, R.\ D.\ 1986, \aj, 92, 1358

\bibitem[Sabbi et~al.(2004)]{SFSR04} Sabbi, E., Ferraro, 
F.~R., Sills, A., \& Rood, R.~T.\ 2004, \apj, 617, 1296

\bibitem[Sills \& Bailyn(1999)]{SB99} Sills, A., \& Bailyn, C.\
D.\ 1999, \apj, 513, 428

\bibitem[Sills \& Lombardi(1997)]{SL97} Sills, A., \& Lombardi, J. C.,
Jr.\  1997, \apjl, 484, L51

\bibitem[Sills et~al.(2000)]{S00} Sills, A., Bailyn, C.\ D., Edmonds,
P.\ D., \& Gilliland, R.\ L.\ 2000, \apj, 535, 298

\bibitem[Stetson(1992)]{stetson} Stetson, P.\ B.\ 1992, ASP 
Conf.\ Ser.\ 25: Astronomical Data Analysis Software and Systems I,
25, 297

\bibitem[Tian et~al. (2006)]{TDHZ06} Tian, B., Deng, L., Han, Z., Zhang, 
X. B. 2006 (astro-ph/0604290)

\bibitem[Warren, Sandquist \& Bolte(2006)]{WSB06} Warren, S. R.,
Sandquist, E. L., \& Bolte, M.\ 2006 (astro-ph/0605047)

\bibitem[Zinn \& Searle(1976)]{ZS76} Zinn, R., \& Searle, L.\ 1976,
\apj, 209, 734

\end{thebibliography}
\end{document}